# SYSTEMATIC REVIEW OF COLLABORATIVE LEARNING ACTIVITIES FOR PROMOTING AI LITERACY

**A Hingle** [a,1]**, A Johri** [b]**,**

[a] George Mason University, Fairfax VA, USA, https://orcid.org/0000-0002-6178-1256
[b] George Mason University, Fairfax VA, USA, https://orcid.org/0000-0001-9018-7574

## ABSTRACT

Improving artificial intelligence (AI) literacy has become an important consideration for academia and industry with the widespread adoption of AI technologies. Collaborative learning (CL) approaches have proven effective for information literacy, and in this study, we investigate the effectiveness of CL in improving AI knowledge and skills. We systematically collected data to create a corpus of nine studies from 2015-2023. We used the Interactive-Constructive-Active-Passive (ICAP) framework to theoretically analyze the CL outcomes for AI literacy reported in each. Findings suggest that CL effectively increases AI literacy across a range of activities, settings, and groups of learners. While most studies occurred in classroom settings, some aimed to broaden participation by involving educators and families or using AI agents to support teamwork. Additionally, we found that instructional activities included all the ICAP modes. We draw implications for future research and teaching.

[1] *A Hingle*
*ahingle2@gmu.edu*

# 1 INTRODUCTION

Artificial intelligence (AI) literacy refers to the knowledge and skills necessary to understand, be empowered to interact with, and make value-informed decisions about AI. Several reviews on AI literacy and competency have explored mechanisms for incorporating AI curriculum in the classroom (Long & Magerko, 2020; Ng et al., 2023; Almatrafi et al., 2024). Educators are designing and implementing various instructional activities to promote AI literacy and make complex AI concepts engaging and accessible for learners of all ages.

Prior work has unequivocally demonstrated the effectiveness of group and team-based collaborative learning (CL) activities for STEM education, including computing and engineering education (Julie et al., 2020; Mercier et al., 2023; Ng et al., 2023). Especially given the effectiveness of CL for computing education more broadly, it is important to understand its efficacy for AI literacy. As yet, there has been little emphasis on systematically understanding the effect of the interaction modality on AI learning (Almatrafi et al., 2024). This paper uses a theory-driven systematic review to examine the effectiveness of using CL to improve AI literacy.

# 2 THEORY GUIDING REVIEW: THE ICAP FRAMEWORK

To systematically and reliably compare the efficacy of CL across studies, we leverage the *Interactive-Constructive-Active-Passive* or *ICAP framework* that distinguishes observable states of learning and thus allows a more effective assessment of educational activities (Chi & Wylie, 2014). According to the ICAP framework, interactive activities that engage learners in instructional settings to co-construct new knowledge yield the best learning outcomes within its hierarchy of the proposed four learning modes. The interactive mode involves developing knowledge beyond the provided materials by collectively adding to understanding by asking questions to elicit further details, giving and receiving feedback, and incorporating other's ideas toward knowledge generation (Chi & Boucher, 2023). The interactive mode is followed by the *constructive* mode, where learners go beyond the initial instructions, and *active* learning activities, where knowledge is applied to similar but non-identical scenarios. *Passive* learning activities like listening or reading are less effective because they require similar cues or contexts for recall (Chi & Wylie, 2014, p.227).

Researchers have found the ICAP framework to be an effective lens to evaluate CL in CS and engineering education (Akgun et al., 2024). Chowdhury et al. (2022) used the ICAP framework to structure graduate-level CS course discussions to foster computational thinking (Chowdhury et al., 2022). Learners collaborated with their peers, allowing them to observe each other's work, seek help, and assess their progress. Facing a challenging activity or struggling, especially for novice learners, encouraged them to switch learning modes from more passive to constructive or active (Chowdhury et al., 2022, p.5). Using the ICAP framework for analysis, we are positioned to understand the types of CL activities implemented and their effectiveness in moving learners across different levels of engagement. It also allows us to compare different studies across stable markers of CL.

The research questions guiding this review are:

RQ1: What are the goals of CL activities for AI literacy, and what CL activities are being implemented?

RQ2: How effective are CL activities in AI literacy development as documented through the ICAP framework?

## 3 REVIEW METHODOLOGY

The Preferred Reporting Items for Systematic Reviews and Meta-Analyses (PRISMA) methodology for conducting reviews was used to coordinate this review (Page et al., 2021).

### 3.1 Database Search and Filtering

We conducted our search and download of potential studies in February 2024. Our search yielded 227 studies. Figure 1 presents a search overview. We describe this process in detail in this section.

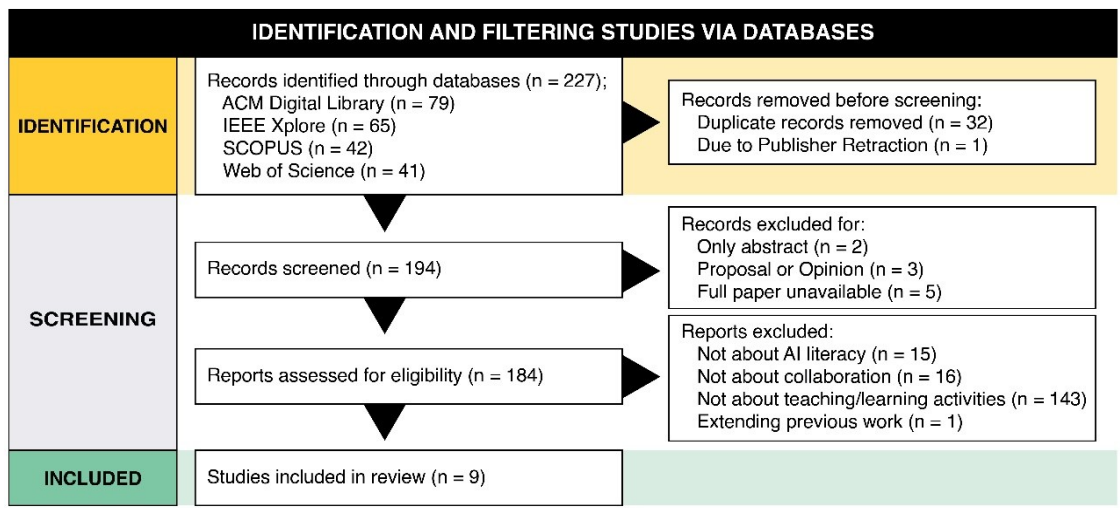

Fig. 1. Search strategy, inclusion, and exclusion criteria following the PRISMA framework.

### Search String

The search strings were constructed to meet four criteria: the research 1) was about AI, 2) described literacy, competency, skills, or knowledge, 3) included a focus on collaborative learning, and 4) presented sufficient details to evaluate the results of an activity. Individual search strings accommodated specific wildcards and other requirements for search across each of the databases - e.g. (ai OR "artificial intelligence") AND (literacy OR competency OR skills OR knowledge OR education) AND (collaboration OR collaborative OR collaborate OR "collaborative learning" OR teams OR groups OR class OR course OR activity) AND ("case study" OR "design study" OR "empirical study" OR intervention). We tested the search strings with the individual criteria to ensure elements were not overlooked.

### Databases

We searched for literature through four databases: 1) ACM Digital Library, 2) IEEE Xplore, 3) SCOPUS, and 4) Web of Science. We also initially examined

ScienceDirect and ProQuest. However, these databases were excluded due to excessive irrelevant studies. Queries with "AI" or "Artificial Intelligence" produced over 10,000 results, and no additional keywords or search strategies effectively reduced the dataset to a manageable size. Alternative filtering methodologies were not employed to maintain reproducibility.

### 3.2 Applying the inclusion and exclusion criteria

We applied the following inclusion and exclusion criteria to refine our search for studies on AI literacy activities published in global conferences or journals:

1. Published between January 2015 and December 2023
2. Published as a conference proceeding or journal article
3. Published in English
4. Focused on AI literacy
5. Focused on collaboration between humans and humans, or humans and AI
6. Presented empirical research (an activity or intervention was described).

We included studies published between 2015 and 2023 because of the significant technological advancements and AI developments during this period. Additionally, although there were few papers on AI literacy before 2015 (Tenório et al., 2023), they often referred to AI in a contextual or theoretical way. Therefore, we limited the scope and relevance of earlier studies. The databases selected include conferences such as CHI, FIE, SIGCSE, and ITiCSE and research papers where relevant engineering and CS education studies have been published. Once the 227 studies were identified, all studies were pre-screened to remove duplicates and any retracted papers. Next, the studies were screened for full paper availability, and any studies mislabelled through the database were removed.

The remaining studies were then downloaded, and two reviewers read through the abstracts of the studies and individually applied the inclusion and exclusion criteria before convening to discuss the results. The reviewers read through the full papers if the abstract was not descriptive about one or multiple inclusion and exclusion criteria elements. After completing this phase individually, they compared the inclusion analysis. The inter-rater agreement for the inclusion and exclusion criteria was 0.9 (Cohen's kappa), which was sufficient for review (Landis & Koch, 1977). All discrepancies were discussed, and the reviewers resolved any inconsistency. Reliability is reported for replicability (McDonald et al., 2019). The most common reasons for exclusion were studies that did not meet all six criteria for inclusion, as this research describes a systematic review of collaborative learning activities for AI literacy. Across the studies that were not included, many papers presented several elements but missed others, especially the requirement for implementing an activity and providing sufficient details to assess its effectiveness.

### 3.3 Analyzing the included studies

Once the final nine studies were identified, two reviewers re-read through the abstract and the full paper individually. Next, they explored the similarities and differences across the work while keeping the ICAP framework and learning modes in mind. The categories included the target audience, skill level, what the CL activity included, and more. They then met and discussed the initial exploration to reach a consistent conceptual framing before they completed the analysis.

*Table 1. Summary of included studies.*

| Reference | Study Outcome | Study Location | Activity Description | Participant Description |
|---|---|---|---|---|
| Druga et al. 2019 | Improving AI Literacy | USA, Germany, Denmark, and Sweden | Workshop with a series of activities | 102 children (ages:7-12 ages) (site: classroom) |
| Ali et al. 2023 | Improving AI Literacy | USA | Card game | 4 grad students and 1 professional (ages: 21-30) (site: classroom) |
| Lee et al. 2022 | Improving AI Literacy | USA | Book Club, Meeting Discussion | 37 educators; 18 middle school teachers and 19 facilitators (site: online community) |
| Long et al. 2022 | Improving AI Literacy | USA | Family group dialogue activities | 13 families; 35 participants (ages: 6-17 and 18+) (site: homes) |
| Sampaio et al. 2023 | Improving ML Literacy & 21st Century Skills | Brazil | Workshop | 32 8 and 9-grade students (12 boys and 20 girls) (site: classroom) |
| Wan et al. 2020 | Improving ML Literacy | USA | Pre-survey, course, discussion | 8 participants (ages: 15-17) (site: classroom) |
| Cai et al. 2019 | Training for Human–AI collaboration | USA | Qualitative laboratory study | 21 pathologists (site: office) |
| He et al. 2023 | Self-perceived Competence | USA | Laboratory setting | 314 participants (site: crowdsourced) |
| De Barros et al. 2023 | Improving Technical AI Skills through PBL | Brazil | Develop ML models | 30 undergraduate computing students (site: classroom) |

## 4 RESULTS AND DISCUSSION

We present the results summarizing the nine selected papers with examples of representative papers to provide context. The final set of studies is listed in Table 1.

### 4.1 RQ1: The Goals and Implementation of CL Activities for AI Literacy

To address RQ1, we present a summary of the studies in Table 1 across the following dimensions: 1) the study goal, 2) the study location, 3) the CL description of the activity, and 4) the study's target population and setting. Overall, in answering RQ1, CL activities have been successfully implemented in various contexts for learners and learning sites. The breadth of potential activities provides a promising outlook for future developments.

## 4.2 RQ2: CL activities for AI literacy learning transition across ICAP modes

To address RQ2, we analyzed the studies using the ICAP framework learning modes (Interactive, Constructive, Active, and Passive) (Chi & Wylie, 2014). Table 2 presents the CL activity description, partners, and the ICAP learning modes.

*Table 2. Analysis through the ICAP Framework.*

| Reference | CL Activity Details | CL Partner(s) | ICAP Framework Modes |
|---|---|---|---|
| Druga et al. 2019 | Playing, sharing & controlling equipment & helping each other with programming. | Student to Student with AI | Passive (watch others play with toys), Active (play with toys), & Constructive (evaluate how to interact with the toys). |
| Ali et al. 2023 | Playing the card game allows players to combine cards creatively, leading to interesting discussions. | Player to Player | Passive (observe how others play card combinations), Constructive (strategize play), & Interactive (respond to challenged plays). |
| Lee et al. 2022 | Educators discuss their goals, resources, & experiences with AI & explore incorporating them into their teaching. | Researcher to Educator to Educator | Passive (listen), Active (read), Constructive (analyze themes in readings), & Interactive (exchange ideas for future use). |
| Long et al. 2022 | Dialogue & interaction between family members while engaging in the three activities. | Family Member to Family Member | Passive (observe others), Constructive (rationalize elements of the activities), & Interactive (agree upon & discuss). |
| Sampaio et al. 2023 | Small groups work with Arduino board, external resistors, & LEDs to build & test circuits to work with the code. | Instructor to Student to Student | Passive (learn concepts & techniques), Active (build example app), & Interactive (debate & evaluate model performance). |
| Wan et al. 2020 | Students learned about *k-means* clustering, discussed how to group clusters of similar faces, & decided on the *k*. | Student to Student working with AI | Passive (class instruction on AI concepts), Active (arrange items in clusters), & Constructive (define groupings of clusters). |
| Cai et al. 2019 | Pathologists reviewed cancer-grade predictions from a model & reflected on their interactions with the model. | Professional working with AI | Passive (observe the results from the model), & Interactive (consult with, critique, & discuss the results). |
| He et al. 2023 | Participants answered a logic problem individually, then had the option to receive "AI advice" & revise the answer. | Individual working with AI | Passive (read the logic problem), Active (select the best answer), Interactive (consult with, critique, & discuss the AI advice). |
| De Barros et al. 2023 | Students in groups of 6 built a predictive model from a dataset using PBL (Agile & Scrum). | Student to Student | Passive (learn concepts), Constructive (design model with a data set), & Interactive (discuss models) |

To explore the modes of the ICAP framework, the classifying verbs associated with each mode were used (Chi & Boucher, 2023) to help guide classification and analysis. All nine studies included multiple modes of collaborative interactions (Figure 2). Every study included components of more individualized learning in the passive and active modes, such as listening to a lecture or reading books or notes, alongside collaborative components, such as playing games with others that required participants to strategize and anticipate moves or working together to find the best model for a dataset. Specifically, nine studies included passive modes, five included active components, seven constructive, and eight interactive.

| STUDIES | ICAP FRAMEWORK MODES | | | |
|---|---|---|---|---|
| | Passive | Active | Constructive | Interactive |
| Druga et al. 2019 | ■ | ■ | ■ | |
| Ali et al. 2023 | ■ | | ■ | ■ |
| Lee et al. 2022 | ■ | ■ | ■ | ■ |
| Long et al., 2022 | ■ | | ■ | ■ |
| Sampaio et al. 2023 | ■ | ■ | | ■ |
| Wan et al. 2020 | ■ | ■ | ■ | |
| Cai et al. 2019 | ■ | | | ■ |
| He et al. 2023 | ■ | ■ | | ■ |
| De Barros et al. 2023 | ■ | | ■ | ■ |

Fig. 2. ICAP framework modes in the included study activities.

This approach of including lower and higher modes aligns with the participant's skill level across the studies, most transitioning from an introductory to an intermediate learning context. Including introductory activities to ensure the learners have opportunities to engage with the theoretical knowledge they need before any synthesis can happen, along with the options to discuss and generate new ideas, is crucial for advancing their comprehension. The most significant leap in learning occurs when moving from the Active to Constructive modes (Chi & Boucher, 2023), equipping learners with the tools and adequately preparing them to create new knowledge essential for fostering deeper learning. All the studies highlight making this transition possible from the structure and design of the intervention.

Overall, in answering RQ2, all the studies that aimed to improve learners' AI literacy (or adjacent constructs) found their specific methodology able to enhance their selected construct. Experimental studies highlighted the need for further research on reliance on AI systems and the effect of AI literacy on these interactions (Cai et al., 2019; He et al., 2023). Likewise, the course-based engagements to improve other domain skills highlighted the necessity of advancing learners' AI literacy to a functional level (De Barros et al., 2023).

## 5  IMPLICATIONS AND FUTURE WORK

### 5.1  Implications for Teaching and Learning

In terms of instruction for improving AI literacy, CL activities should be incorporated within teaching to allow learners to co-create knowledge contextualized to their needs. Depending on the goals of instruction, activities can target literacy, specific use-case knowledge, or domain knowledge. The studies presented in this paper are examples of how CL can be incorporated with these contexts in mind. The modality of the activity is an essential consideration for educators developing AI literacy programs. Special attention should be given to how learners contextualize AI in their micro and macro decision-making. Furthermore, it may be necessary to integrate CL with other activities and assessments to enhance comprehension and engagement.

## 5.2 Implications for Research

There are several implications for future research. Although there is evidence that CL supports AI literacy, as seen through the ICAP framework, more studies are needed across different settings, and more design-based research within a setting is required to understand if multiple iterations can improve the outcomes. Additionally, although the review includes studies implemented in several countries, there is a disproportionate representation of work from the USA. This does not mean other countries are not engaging in AI literacy efforts, but the implementation context may differ. AI literacy does not need to take on a singular form worldwide but rather be contextual to the needs of the specific communities where it is implemented. In some instances, AI literacy might be integrated with other literacies, such as data or digital literacy, which alters the framing of these efforts (Julie et al., 2020). Evidence for the use of AI partners is present, but more work is needed to ascertain how computing agents can best support the learning of AI itself. Studies on human-AI teaming and reliance are on the rise (Zhang et al., 2021; Zhou & Chen, 2019), and understanding how CL interactions between human-human and human-AI affect learners' understanding of the capabilities and constraints of AI is important.

In pursuit of advancing collaborative AI literacy work, it is essential to explore other established theories, methodologies, and activities, just as these studies do, building on the substantial existing body of research on collaboration. Integrating this work into contexts that resonate most with learners' needs while highlighting methods that instructors are well-versed in will be the key to successful implementation.

## 6 LIMITATIONS

This study has several limitations that are in common with systematic reviews. Firstly, the final nine studies identified are implemented in different learning contexts, learners, and target audiences. Therefore, it is challenging to summarize findings across the studies, and this should be done with care. Secondly, due to the abundance of studies being published using keywords such as AI and collaboration and others that use different descriptors, though referring to the same constructs, our final dataset may potentially exclude relevant studies. Still, the search string, inclusion and exclusion criteria, and our review of the studies represent studies that meet the criteria set out in the research questions. Additionally, although the collaborative processes described in the studies are focused on the learner, other elements could be missing from the context. Collaboration in real-world settings often includes interactions with colleagues, artifacts such as company repositories, or other resources (Cai et al., 2019), which can provide guidance that is difficult to replicate in a controlled learning environment.

## 7 ACKNOWLEDGEMENTS

This work is partly supported by US NSF Awards 2319137, 1954556, and USDA/NIFA Award 2021-67021-35329. Any opinions, findings, conclusions, or recommendations expressed in this material are those of the authors and do not necessarily reflect the views of the funding agencies. The material in this paper is part of the dissertation of the primary author.